\documentstyle[revtex]{aps}
\begin{document}
\def\overlay#1#2{\setbox0=\hbox{#1}\setbox1=\hbox to \wd0{\hss #2\hss}#1%
\hskip -2\wd0\copy1}
\begin{title}
New Fermionic Description of Quantum S = 1/2 Antiferromagnet
\end{title}
\author{A. M. Tsvelik \cite{ucsd}}
\begin{instit}
Department of  Physics, Harvard University, Cambridge, Massachusetts 02138
\end{instit}
\begin{abstract}
 A novel approach to S = 1/2 antiferromagnets with strong quantum
 fluctuations based on the representation of spin-1/2 operators as
 bylinear forms of real
 (Majorana) fermions is suggested.
 This representation   has the advantage of being irreducible
without  any constraints on the fermionic  Hilbert space. This property
 allowed me to write an effective theory for low-lying excitations in a
 spin liquid state of S = 1/2 antiferromagnet. It is  proven that these
 excitations are S = 1 real (Majorana) fermions.



\end{abstract}

  The traditional picture representing magnets as arrays of weakly interacting
 rigid rotors fails to describe systems with strong quantum fluctuations. Such
fluctuations can be increased by a small value of spin, a frustrated
interaction,
or by a large value of the rank of a spin group. An alternative point of view
on strongly fluctuating systems  has been gradually emerging from
the original Anderson's conjecture of a gapless "spin liquid" state in two-
dimensional S = 1/2 Heisenberg systems [1]. Later he used these ideas  to
explain unusual magnetic properties of the copper oxide superconductors [2]
which
 has attracted much attention. The Anderson's proposal for systems
 with strong fluctuations is to concentrate on bonds between spins rather than
on spins themselves and thus define  slow bond variables $\Delta(r,r')$
(usually
 called RVB -
resonating valence bond -  order parameter). This approach assumes that  a
theory
 of spin liquid is a lattice gauge theory.

  Current efforts to develop a gauge invariant theory of spin liquid are based
on
the Wigner and Schwinger [3] representation of spin operators:
\begin{eqnarray}
S^a = b^+_{\alpha}\sigma^a_{\alpha,\beta}b_{\beta}\\
b^+_1b_1 + b^+_2b_2 = 2S
\end{eqnarray}
where $\sigma^a$ are the Pauli matrices and $b^+,b$ are either Bose or
Fermi operators.

  The above representation is used as a basis for a mean field theory where
the Heisenberg exchange term is decoupled with an auxilary field $\Delta(r,r')$
by the Hubbard-Stratonovich transformation and the constraint (2) is enforced
by a local Lagrange multiplier field $\lambda(t,r)$. Depending on the
 expected answer some authors use the bosonic representation (see, for
example, Refs.(4,5)) and some prefer the fermionic one (see Refs.(6-9)).
 The resulting mean field theory describes an incompressible liquid of
 bosons or fermions.  The usual argument is that the  ambiguity of
 statistics is irrelevant due to the incompressibility of the liquid
 (the constraint (2)).  Meanwhile nobody has succeded in taking the
 incompressibility into account.
 The unresolved constraint (2) reveals itself as gauge long distant forces
 between  low-lying excitations and possibly to their confinement
(see the corresponding expressions in Refs.(6-8)). The latter would mean
that conventional spin wave behavior is restored.  Thus the theory  in the
 continuous limit
still contains strong interactions  which is unsatisfactory.

 In this Letter I suggest another approach based on  the special
 representation of
 spin-1/2 operators which does not contain constraints. This representation
 is well
known in the high energy physics;
the spin-1/2 operators are represented as bylinear forms of real (Majorana)
 fermions
transforming according to the adjoint representation of the SU(2) group.
I  begin the discussion from a brief
description of the Majorana fermions.

  Let us consider lattice fermions $\eta_a(r)$ (a = 1,2,3; $r$ labels
 lattice sites) with the following commutation relations:
\begin{equation}
[\eta_a(r),\eta_b(r')]_+ = \delta_{a,b}\delta_{r,r'}
\end{equation}
 In mathematics the algebra (3) is called the Clifford algebra. For
 the particular
 case of the lattice with only one site $\eta_a$ coincide with the spin-1/2
 matrices. In general the commutation
relations (3) can be obtained from a quantization of the following Lagrangian:
\begin{equation}
L_0 = \sum_r i\eta_a(r,t)\partial_t\eta_a(r,t)
\end{equation}

  It is easy to check that the following representation reproduces the
 commutation  relations of the spin operators:
\begin{equation}
S^a(r) = -i\epsilon_{abc}\eta_b(r)\eta_c(r)
\end{equation}
 It follows from (3) that $\vec S^2 = 3/4$, i.e. the representation (5) is
 irreducible representation of  S = 1/2.

 The irreducibility of the representation (5) gives us an advantage of an
 unconstrainted Hilbert space.

 As the Schwinger representation (1,2) the representation (5)  has
a local $Z_2$ gauge symmetry: it is invariant with respect to the
transformation
\begin{equation}
\eta_a(r) \rightarrow (-1)^{q(r)}\eta_a(r)
\end{equation}
($q(r) = \pm 1$)
 Therefore  it is $2^N$-valued ($N$ is a number of lattice sites) and the
number
 of states of the fermions is larger that the number of spin states.

  Now let us consider a Heisenberg antiferromagnet with S = 1/2 which
 Hamiltonian is given by
\begin{equation}
H = \sum J(r,r')S^a(r)S^a(r') - h^aS^a(r)
\end{equation}
I use the representation (5); than the corresponding Lagragian is
 $L = L_0 - H$. After a decoupling of the Heisenberg exchange interaction
 with an auxilary field
 $\Delta(r,r')$ this Lagrangian acquires the following form:
\begin{eqnarray}
L = \sum_r\left[ i\eta^a(r)\partial_t\eta^a(r) +
 \Delta(r,r')\eta^a(r)\eta^a(r') +
ih^a\epsilon_{abc}\eta^b(r)\eta^c(r) +\frac{\Delta^2(r,r')}{2J(r,r')}\right]
\end{eqnarray}

 The Lagrangian (8) is invariant under the gauge transformation of fermions
 (6) and the order parameter field $\Delta(r,r')$:
\begin{equation}
\Delta(r,r')\rightarrow \Delta(r,r')(-1)^{[q(r) - q(r')]}
\end{equation}
and also under transformations from the crystal lattice group:
\begin{equation}
\Delta\rightarrow C^{-1}\Delta C
\end{equation}
  In the mean field approximation  $\Delta(r,r')$ is treated as a
 time-independent field  determined self-consistently to get a state with
 minimal energy. The parameter of the expansion around the mean field
 saddle point is $1/n$ where $n = 3$ is a
number of fermionic "colors". In principle, one can generalize this
 approach to the case where  the spin operators are substituted by
 generators of the $O(n)$  symmetry group thus getting a formal parameter.

  Stability of the above fixed point depends on a particular choise of
$J(r,r')$.
I will not discuss this question and instead concentrate on general properties
of
the solution provided it is stable.

  In the absence of a spin long range order one can characterize the ground
 state by expectation values of two gauge invariant operators:
\begin{eqnarray}
\epsilon(r,r') = <\Delta^2(r,r')>\\
W(C) = sign<\Pi_C \Delta(r,r')>
\end{eqnarray}
 where the product in Eq.(12) is taken along a contour $C$.

  Interesting situations arise only if a spacial variation of $|\Delta(r,r')|$
 is small. Otherwise we will have trivial dimer (or spin-Peierls) states with
 a short range order.  Neglecting  fluctuations
of  $|\Delta|$ and substituting it for a coordinate independent quantity
$\Delta_0$
 found from the condition
of minimal free energy we get from Eq.(8) the following effective Lagrangian
for
 low energy modes of the theory:
\begin{equation}
L = \sum_ri\eta_a(r)\partial_t\eta_a(r) + ih^a\epsilon_{abc}\eta^b(r)\eta^c(r)
+
\sum_{<r,r'>}i\Delta_0\eta_a(r)\sigma(r,r')\eta_a(r')
\end{equation}
where $\sigma(r,r') = \pm 1$ is a  $Z_2$ gauge field defined on the links of
the
 lattice with nonvanishing $J(r,r')$.

 The Lagrangian (13) allows a further  simplification  if in  the ground state
  $W(C) = 1$ for all loops (uniform states).  The uniform state could be stable
 provided there are no flat parts on the Fermi surface. Usually such parts are
 removed by interactions with next nearest neighbours.

  For the further analysis  it is convenient to make a Fourier transformation
of the fermionic fields. From Eq.(3) it follows that the transformed fields
 satisfy  the following commutation relations:

\begin{equation}
[\eta^a(\vec k),\eta^b(\vec k')] = \delta_{ab}\delta_{\vec k + \vec k',0}
\end{equation}

  Then from Eq.(8) I derive the Hamiltonian for the uniform state (the Zeeman
 term is omitted):
\begin{equation}
H = \sum_{\vec k} \epsilon(\vec k)\eta^a(-\vec k)\eta^a(\vec k) +
 \sum_{\vec q,\vec p,\vec k}
J(\vec p - \vec k):\eta^a(\vec k)\eta^a(-\vec k +
 \vec q)::\eta^b(-\vec p)\eta^b(\vec p - \vec q):
\end{equation}
where the dots denote the normal ordering  and
\begin{equation}
 \epsilon(\vec k) = <\eta^a(-\vec k)\eta^a(\vec k)>\sum_{<r,r'>}J(r,r')
\cos{[\vec k(\vec r - \vec r')]}
\end{equation}

  The energy $\epsilon(\vec k)$ vanishes on some surface; since according to
 Eq.(16)
$\epsilon(\vec k) = - \epsilon(-\vec k)$ this surface  divides the Brillouin
 zone
into two equal parts $\Omega_+$ (where $\epsilon > 0$) and $\Omega_-$
 ($\epsilon < 0$).
 Then $\eta(\vec k)$ and $\eta(-\vec k)$ with $\vec k$ belonging to
 $\Omega_-$ are
 creation and annihilation  operators which returns us to the conventional
 Fermi liquid
picture. Fermions fill a half of the Brillouin zone which garantees that
 the spin sum rules are satisfied. In fact, the uniform state is
 a chargeless Fermi liquid. It follows from the fact that due to a nonsingular
 character of the  interaction in Eq.(15)
all scattering effects become irrelevant close to the Fermi surface. Therefore
  fermions with small energies propagate coherently. Thus this state of a
 quantum antiferromagnet does  have fermions as elementary excitations, but
 these fermions have spin 1 instead of the usual 1/2
since they belong to the adjoint representation of the SU(2) group!
 As in a conventional Fermi liquid
 the system  has a linear specific heat $C_v \propto T$;
 the magnetic susceptibility $\chi(\omega,\vec q)$ is almost
 $\vec q$-independent at small $\vec q$, has a weak singularity at
 $|\vec q| = 2p_F$ and exhibits a
significant  imaginary part in the broad range of frequences  $\omega \sim J$
 without an energy gap.  Such behavior
 has been  observed in various strongly fluctuated magnetic systems.

  Except of the uniform sates another possibility exists which arises when
 the mean field ground state favors $W(C) = -1$ for certain loops. Such
 ground states has been intensively
 discussed in the literature as flux (or chiral)
 states (see Refs.(6-10)). The present case differs from the conventional
 formulation of the problem in two aspects:
 the effective hopping integrals $\Delta(r,r')$ are now real
and the fermionic filling factor is always equal to 1/2.

A  flux state is  periodic, but
its periodicity does not coincide with the initial periodicity of the lattice.
 A new
elementary cell arises which is defined as a minimal cell surrounded by
 contours with  $W(C) = 1$.
Since the initial symmetry is broken  a flux phase  can arise at low
 temperatures only via a phase
transition. Apparently in the high temperature phase the fermions are
confined into ordinary spins and  the transition release them. Since the
broken symmetry is discrete, such transition may happen even in two
dimensions. Whether it happen or not  is not  clear, however.
 For the lattice $Z_2$ gauge theory which is similar to the theory I
discuss, it was established in Ref.11 that there is no phase transition
 in d = 2 and it occurs only in d = 3 as  a second order transition.
Therefore the problem of existence  of two dimensional flux phase
of a  quantum antiferromagnet requires further studies.
As far as hipothetical three dimensional  flux states are concerned they
  would have the following properties.  As we know, the spectrum of fermions
 in a flux state is  either gapful or has
  conical singularities: $\epsilon(\vec k) \propto |\vec k - \vec k_0|$ [10].
 The specific heat in the latter  case looks like for an ordinary
 antiferromagnet: $C_v \propto T^3$, but the
magnetic properties are essentially different. In particular, the magnetic
 susceptibility
 behaves as $\chi\propto h^{2}$. It cannot be excluded that the remaining
 interactions can cause singularities in the
nonlinear magnetic susceptibility. I emphasise that elementary excitations
 are again S = 1 fermions and not spinons as it follows from the ground
 state wave function  postulated in Ref.10.

  Unfortunately, I have not managed to generalize the described approach
 to the $SU(2/1)$ group
which is necessary if one wants to include effects of doping.
It prevents me from a straightforward  application of this technique
to the Hubbard or the $t-J$ models. At the present time the described procedure
can be used only  for problems with well defined spin operators,
 for example, for the Kondo lattice problem [12].

 I am greatful to P. W. Anderson and P. Coleman for intensive and valuable
 discussions and to
P. B. Wiegmann for sending me his preprint. This work was supported by the
 National Science Grant Foundation, through grant DMR 91-15491 and through
 the Harvard University Materials Research Laboratory.

\end{document}